\documentclass[iop]{emulateapj}
\usepackage{natbib}
\usepackage{amsmath}
\usepackage{graphicx}
\usepackage{hyperref}

\shorttitle{Dark Matter in Clusters from Radio Emission}

\shortauthors{Storm, Jeltema, Profumo, \& Rudnick}

\begin{document}

\title{Constraints on Dark Matter Annihilation in Clusters of Galaxies from Diffuse Radio Emission}

\author{Emma Storm\altaffilmark{1}, Tesla E. Jeltema\altaffilmark{1,2}, Stefano Profumo\altaffilmark{1,2}, Lawrence Rudnick\altaffilmark{3}}

\altaffiltext{1}{Department of Physics,  University of California, 1156 High St., Santa Cruz, CA 95064, USA}
\altaffiltext{2}{Santa Cruz Institute for Particle Physics,  University of California, 1156 High St., Santa Cruz, CA 95064, USA}
\altaffiltext{3}{Minnesota Institute for Astrophysics, School of Physics and Astronomy, University of Minnesota, 116 Church Street SE, Minneapolis, MN 55455, USA}

\begin{abstract}
Annihilation of dark matter can result in the production of stable Standard Model particles including electrons and positrons that, in the presence of magnetic fields, lose energy via synchrotron radiation, observable as radio emission. Galaxy clusters are excellent targets to search for or to constrain the rate of dark matter annihilation, as they are both massive and dark matter dominated. In this study, we place limits on dark matter annihilation in a sample of nearby clusters using upper limits on the diffuse radio emission, low levels of observed diffuse emission, or detections of radio mini-haloes. We find that the strongest limits on the annihilation cross section are better than limits derived from the non-detection of clusters in the gamma-ray band by a factor of $\sim3$ or more when the same annihilation channel and subtructure model, but different best-case clusters, are compared. The limits on the cross section depend on the assumed amount of substructure, varying by as much as 2 orders of magnitude for increasingly optimistic substructure models as compared to a smooth NFW profile. In our most optimistic case, using the results of the Phoenix Project \citep{Gao2012phx}, we find that the derived limits reach below the thermal relic cross section of $3\times 10^{-26}$~cm$^3$~s$^{-1}$ for dark matter masses as large as $400$~GeV, for the $b\overline{b}$ annihilation channel. We discuss uncertainties due to the limited available data on the magnetic field structure of individual clusters. We also report the discovery of diffuse radio emission from the central $30$-$40$~kpc regions of the groups M49 and NGC4636. 
\end{abstract}


\section{INTRODUCTION}

Clusters of galaxies are the most massive virialized objects in the universe. About $80\%$ of the mass of clusters is comprised of dark matter, making them good candidates for astrophysical searches for a signature from particle dark matter. Among the best motivated particle candidates for dark matter are weakly interacting massive particles, or ``WIMPs''. WIMPs can self-annihilate to Standard Model particles, including electrons and positrons. Example particle theories that predict WIMP dark matter are supersymmetric extensions to the Standard Model (for a review, see e.g., \citealt{Jungman1996}) and universal extra dimensions (for a review, see e.g., \citealt{Hooper2007}). The products of WIMP annihilations generically yield a broad spectrum of electromagnetic emission in cluster environments that is potentially observable across a wide range of frequencies, from radio to gamma rays \citep{Colafrancesco2006}. 

Recent searches for dark matter annihilation have largely focused on indirect detection through gamma-ray emission. Gamma rays are primarily produced in WIMP annihilation by two classes of processes:

(i) in the {\em prompt} emission from the two-photon decay of neutral pions produced in the hadronization of strongly-interacting particles produced in the WIMP annihilation, or from radiative processes such as internal bremsstrahlung, and

(ii) in the {\em secondary} emission when high-energy electrons and positrons (either directly produced or resulting from the decays of particles produced in the annihilation process, e.g. charged pions) up-scatter ambient photons to gamma-ray energies, a process known as Inverse Compton (IC) scattering.

An additional secondary gamma-ray emission process, typically subdominant, is associated with secondary bremsstrahlung off of ambient gas of the same electrons and positrons responsible for the IC emission \citep{Colafrancesco2006}.

Galaxy clusters have not yet been detected in gamma rays (see recently, e.g., \citealt{Macias-Ramirez2012,Han2012a}). Upper limits on the gamma ray emission have been used to place constraints on the dark matter annihilation cross section. Previous studies that placed constraints on dark matter annihilation or decay in clusters have focused on searching for gamma-ray emission using the Large Area Telescope (LAT) onboard the \textit{Fermi Gamma Ray Space Telescope} at the GeV scale \citep{Ackermann2010dm,Dugger2010,Ando2012,Huang2012,Nezri2012,Han2012a}, and using ground-based Cherenkov telescopes, including H.E.S.S. \citep{Abramowski2012} and MAGIC \citep{Aleksic2010} at the TeV scale. 

The leptons that IC-upscatter photons to GeV and TeV energies also lose energy via synchrotron radiation in the presence of magnetic fields, producing emission that is potentially observable at radio frequencies. Diffuse radio emission from the intracluster medium (ICM) in the form of approximately spherically symmetric haloes and mini-haloes has been observed from more than 40 clusters to date, indicating the existence of both a relativistic population of electrons/positrons and large scale magnetic fields at the $\sim\mu$G level (for a review see e.g., \citealt{Feretti2012}). Giant radio haloes are typically $\sim1$ Mpc in size, tend to follow the distribution of the thermal X-ray emission and are therefore roughly spherical, and are characterized by steep energy spectra and low surface brightness \citep[e.g.,][]{Thierbach2003}. The first discovered, and perhaps best studied member of this class is the halo of the Coma Cluster \citep[see e.g.][]{Large1959,Willson1970,Kim1990,Kronberg2007,Brown2011a}. Mini-haloes are similar in shape to haloes but are on the order of hundreds of kiloparsecs in size and are found in clusters with cool cores, which are characterized primarily by a very short central cooling time ($\ll1$~Gyr), and additionally show a sharp increase in X-ray surface brightness and a corresponding decrease in temperature at the center of the cluster \citep[e.g.,][]{Hudson2010}. The prototype of this class is the mini-halo found in the Perseus Cluster \citep{Burns1992}.

The origin of these relativistic particles responsible for the diffuse radio emission is, however, at present unclear. They perhaps were injected into the ICM by AGN or supernova activity within cluster galaxies and then reaccelerated by turbulence in the ICM caused by mergers \citep[e.g.,][]{Brunetti2001,Petrosian2001,Fujita2003,Cassano2005}. They could also be the products of cosmic ray hadron collisions with particles in the ICM \citep[e.g.,][]{Dennison1980,Blasi1999,Pfrommer2004}. This scenario requires that magnetic fields be strong, at least comparable to the strengths inferred from Faraday Rotation Measures (RMs), so that the gamma ray fluxes from cosmic rays do not exceed current limits \citep{Jeltema2011,Ackermann2010ul}. Finally, the relativistic electrons and positrons could also be the result of dark matter annihilation \citep{Colafrancesco2006,Perez-Torres2009}.

Most clusters show no or only low levels of detectable diffuse radio emission. However, \citet{Brown2011b} stacked a sample of 105 clusters with no detected radio haloes and found a $6\sigma$ detection of radio emission, which suggests that even clusters with no currently observed radio emission host, at some level, both relativistic populations of electrons and positrons and large-scale, $\mu$G magnetic fields. The relativistic electrons and positrons potentially produced by dark matter annihilation in clusters would synchrotron radiate away their energy in the presence of cluster-strength magnetic fields. We can thus use the radio upper limits or low levels of radio emission from clusters to constrain the rate of dark matter annihilation. In this paper, we place limits on the dark matter annihilation cross section using primarily upper limits on or low levels of the diffuse radio emission from a selection of nearby clusters and also using some clusters with detected mini-haloes. Our study greatly improves contraints on dark matter with radio observations over previous studies by carefully selecting optimal targets with little or no observed diffuse radio emission, as opposed to clusters with bright radio halos that have been considered so far \citep[see e.g.][]{Colafrancesco2006}.

The paper is organized as follows. In Sec.~\ref{sec:dm}, we describe our models for the dark matter and magnetic field profiles in clusters. In Sec.~\ref{sec:selection} we present the sample of clusters and radio data we employ for our analysis, report on the observation of diffuse emission from M49 and NGC4636, and discuss the impact of studies of intracluster magnetic fields on our sample. Sec.~\ref{sec:results} presents limits on the dark matter annihilation cross section and discusses the implications and caveats of our results, including comparisons to previous studies and effects of uncertainties in the assumed dark matter profiles and magnetic field structure. We conclude in Sec.~\ref{sec:end}. Throughout this paper, we assume a $\Lambda$CDM cosmology with $H_0=100h$~km~s$^{-1}$~Mpc$^{-1}$, $h=0.70$, $\Omega_{m}=0.27$, and $\Omega_{\Lambda}=0.73$.

\section{DARK MATTER AND MAGNETIC FIELD MODELING}\label{sec:dm}

The spectral flux density due to dark matter annihilation is:
\begin{equation}\label{eq:dmflux}
  S_{\nu} = \frac{\langle \sigma v \rangle}{8\pi m_{\chi}^2}\left(E\frac{dN_{\nu}}{dE}\right)J,
\end{equation}
where $\langle \sigma v \rangle$ is the thermally-averaged zero-temperature dark matter annihilation cross section times velocity, $m_{\chi}$ is the dark matter particle mass, and $EdN_{\nu}/dE$ is the synchrotron energy spectrum per frequency per dark matter annihilation event. The observable flux density $S_{\nu}$ is typically measured in Jansky (Jy) in the radio. $J$ is the line-of-sight integral of the dark matter density squared, integrated over the angular size of the emission region $\Delta \Omega$:
\begin{equation}\label{eq:Jfact}
  J = \int_{\Delta \Omega} \mathrm{d} \Omega \int_{l.o.s.}\rho_{\chi}^2(l) \mathrm{d}l
\end{equation}
We take the size of the emission region to match that of the available radio data, either the size of the observed diffuse emission or the size used to place upper limits. We use a number of different models for the dark matter density profile, described in the following sections.

\subsection{NFW Density Profile}

In our analysis we use four models for the dark matter density profile. The most conservative, with no substructure, is the Navarro-Frenk-White (NFW) profile \citep{Navarro1996, Navarro1997}:
\begin{equation}\label{eq:rhoNFW}
  \rho_{NFW}(r) = \frac{\rho_s}{\frac{r}{r_s}\left(1+\frac{r}{r_s}\right)^2},
\end{equation}
where the central density $\rho_s$ and the scale radius $r_s$ are determined by observations. Following the derivation in \citet{Ackermann2010dm}, we use the scaling relationship derived from X-ray observations of clusters from \citet{Buote2007} to determine $r_s$ and $\rho_s$ from the virial mass and radius, $M_{vir}$ and $r_{vir}$:
\begin{equation}\label{eq:cM}
  c_{vir} = 9\left(\frac{M_{vir}}{10^{14}h^{-1}}\right)^{-0.172},
\end{equation}
where the concentration $c_{vir} = r_{vir}/r_s$. The virial overdensity $\Delta_{vir}\approx 98$ is defined with respect to the critical density of the universe \citep{Bryan1998}, with $M_{vir}=\frac{4\pi}{3}\Delta_{vir}\rho_c r_{vir}^3$. We determine $M_{vir}$ and $r_{vir}$ from $M_{500}$, where the overdensity is 500 times the critical density, obtained from X-ray observations \citep{Chen2007}. We correct $M_{500}$ for our current cosmology and reduce it by the gas fraction $f_{gas}$, also reported in \citet{Chen2007}, so that we are left with only the mass of the dark matter in the cluster: $M_{500}~\rightarrow~M_{500}(1-f_{gas})h$. We use the corrected $M_{500}$ to determine $r_{500}$ and use these parameters to determine $M_{vir}$, $r_{vir}$, and  $r_s$ using the following equations, derived in Appendix C of \citet{Hu2003}:
\begin{equation}\label{eq:Mdel}
  \frac{M_{500}}{M_{vir}} = \frac{\Delta_{500}}{\Delta_{vir}} \left(\frac{r_{500}}{r_{vir}}\right)^3
\end{equation}
and
\begin{equation}\label{eq:rsrv}
  \Delta_{500} f(r_s/r_{vir}) = \Delta_{vir} f(r_s/r_{500}),
\end{equation}
where
\begin{equation}\label{eq:NFWint}
  f(x) = x^3[\mathrm{ln}(1+x^{-1})-(1+x)^{-1}].
\end{equation}
We use $M_{vir}$, $r_{vir}$, and $r_s$ to solve for $\rho_s$ \citep{Hu2003}:
\begin{equation}\label{eq:rhos}
  \rho_s = \frac{M_{vir}}{4 \pi f(r_s/r_{vir})}
\end{equation}
\subsection{Effects of Substructure}

In cold dark matter (CDM) cosmologies, dark matter haloes are structured hierarchically. Clusters are observed to host subhaloes down to at least the scale of dwarf galaxies, approximately $10^7~M_{\odot}$, and are predicted to contain substructure down to $\sim10^{-6}~M_{\odot}$ (see \citet{Green2005}, but also \citet{Profumo2006}). Since the $J$ factor for dark matter annihilation is proportional to the density squared, the amount of substructure is critical, and can increase $J$ by one or more orders of magnitude compared to a smooth NFW density profile. In clusters, tidal stripping tends to destroy subhaloes near the cluster center, so substructure is preferentially found towards the cluster outskirts \citep{Gao2012phx}. Generally the density profile of subhaloes is more extended than the smooth central halo, as seen in simulations \citep{Nagai2005,Gao2012phx}.

We consider here multiple substructure models. Two of these models are also used in \citet{Ackermann2010dm}: a conservative model, with a subhalo cutoff mass of $10^7~M_{\odot}$ and a fraction of the total halo mass in substructure, $f_s$, of $10\%$, and an optimistic model, with a cutoff mass of $10^{-6}~M_{\odot}$ and $f_s$=0.2. We use the formalism in \citet{Colafrancesco2006} to determine the dark matter density profiles $\rho_{CON}$ and $\rho_{OPT}$. In the notation of this framework, the conservative setup corresponds to a substructure contrast factor of $\Delta^2=1.3\times10^5$, and the optimistic setup corresponds to $\Delta^2=3.0\times10^5$.

We also consider a third model based on the results of the Phoenix Project, which is a series of dark matter simulations following the evolution of cluster-sized haloes \citep{Gao2012,Gao2012phx}. These simulations adopt the cosmological parameters of the Millenium Simulation \citep{Springel2005}, which are now known to be inconsistent with current cosmological parameters. However, the main difference is the value of $\sigma_8$, which affects the number of clusters in a universe-sized simulation but not the properties of individual cluster halos. This model adopts a cutoff subhalo mass of $10^{-6}~M_{\odot}$, resulting in a subhalo mass fraction of approximately $27\%$, and is more extended than either of the previous models.

The Phoenix simulations use $M_{200}$ and $r_{200}$ to express their results in terms of a boost factor and substructure surface brightness \citep{Gao2012}:
\begin{equation}
  b(M_{200})=1.6\times10^{-3}\left(\frac{M_{200}}{M_{\odot}}\right)^{0.39}
\end{equation}
\begin{equation}\label{eq:PHXflux}
  S_{sub}(r) = \frac{16b(M_{200})L_{main}}{\pi \mathrm{ln}(17)}\frac{1}{r^2+16r_{200}^2},
\end{equation}
where $L_{main}$ is the luminosity of the smooth halo, which is well-described by an NFW profile. We use Eqns.~(\ref{eq:Mdel}) and~(\ref{eq:rsrv}) to solve for $M_{200}$ and $r_{200}$, defined through the overdensity $\Delta_{200}=200\rho_c$, from $M_{500}$ and $r_{500}$. We then translate Eqn.~(\ref{eq:PHXflux}) to a $J$-factor ``surface brightness'', following \citet{Han2012a}. Thus the resulting $J$ factor for the Phoenix simulations is:
\begin{equation}\label{eq:Jphx}
  J_{PHX} = J_{NFW} + J_{sub}
\end{equation}
where
\begin{equation}
  J_{sub} = \int_{\Delta\Omega}\mathrm{d}\Omega\int \frac{16b(M_{200})J_{NFW}}{\pi ln(17)}\frac{2 \pi r \mathrm{d}r }{r^2+16r_{200}^2}
\end{equation}
\subsection{Electron/Positron Signal from Dark Matter Annihilation}
In order to characterize the synchrotron spectrum, we need the equilibrium electron and positron spectra, which result by solving the full diffusion equation:
\begin{equation}\label{eq:diff}
  \begin{split}
    \frac{\partial}{\partial t} \frac{dn_e}{dE} &= \nabla \left[D(E,{\bf x})\nabla \frac{dn_e}{dE} \right]\\
    &+ \frac{\partial}{\partial E}\left[b_{loss}(E,{\bf x})\frac{dn_e}{dE} \right] + Q(E,{\bf x})
  \end{split}
\end{equation}
where $dn_e/dE$ is the equilibrium electron or positron density spectrum, $Q(E,{\bf x})$ is the source term, $D(E,{\bf x})$ is the spatial diffusion coefficient, and $b_{loss}(E,{\bf x})$ is the energy loss term described below. In clusters of galaxies, the timescale on which electrons radiate is much shorter than the spatial diffusion timescale, so we can neglect the time dependence on the l.h.s. as well as the spatial dependence on the r.h.s. of Eq.~(\ref{eq:diff}) (see e.g., Appendix A of \citealt{Colafrancesco2006} for a discussion of the role of diffusion in clusters). The source term is proportional to the injected spectrum of electrons/positrons per dark matter annihilation. The expression for the equilibrium density spectrum is:
\begin{equation}
\frac{dn_e}{dE} = \frac{\langle \sigma v \rangle \rho_{\chi}^2}{2 m_{\chi}^2 b_{loss}(E)}\int_E^{m_{\chi}} \mathrm{d}E' \frac{dN_{e,inj}}{dE'}
\end{equation}
We use the DMFIT package \citep{Jeltema2008} (in turn derived from the DarkSUSY package \citep{Gondolo2004}) to calculate the electron/positron injection spectra per dark matter annihilation. The energy loss term, in the low redshift limit, is the sum of synchrotron, IC, bremsstrahlung, and Coulomb losses:
\begin{equation}
  \begin{split}
    b_{loss}(E) &= b_{syn} + b_{IC} + b_{brem} + b_{coul} \\
    &\approx0.0254\left(\frac{E}{1\mathrm{GeV}}\right)^2\left(\frac{B}{1 \mu \mathrm{G}}\right)^2\\
    &+ 0.25\left(\frac{E}{1\mathrm{GeV}}\right)^2\\
    &+ 1.51n(0.36+\mathrm{log}(\gamma/n))\\
    &+ 6.13(1+\mathrm{log}(\gamma/n)/0.75)
  \end{split}
\end{equation}
The energy loss term $b_{loss}(E)$ has units of $1\times10^{-16}$~GeV s$^{-1}$, where $n$ is the average thermal electron density, $\approx1\times 10^{-3}$~cm$^{-3}$ for all clusters. For GeV electrons and positrons, synchrotron and IC losses dominate; when $B > B_{CMB}\approx 3~\mu$G, synchrotron losses dominate over IC losses.

\subsection{Synchrotron Emission}

In clusters, relativistic electrons radiate their energy via synchrotron emission in the presence of a magnetic field $B(r)$, and live in a background plasma with electron density $n(r)$ and plasma frequency $\nu_p = 8890[n(r)/1$~cm$^{-3}]^{1/2}$~Hz. The power per frequency of emitted synchrotron radiation for a single electron (or positron) with energy $E=\gamma m_ec^2$, averaged over all incoming directions is \citep{Longair2011}:
\begin{equation}\label{eq:Pelec}
  P_{\nu}(\nu,E) = \int_0^{\pi}\mathrm{d}\theta \frac{\mathrm{sin}\theta}{2} 2\pi \sqrt{3} r_0 m_ec \nu_0 \mathrm{sin}\theta F\left(\frac{x}{\mathrm{sin}\theta}\right) 
\end{equation}
where $r_0=e^2/(mc^2)$ is the classical electron radius, $\theta$ is the pitch angle, $\nu_0 = eB/(2\pi mc)$ is the nonrelativistic gyrofrequency. The quantities $x$ and $F$ are defined as follows:
\begin{equation}\label{eq:x}
x \equiv \frac{2\nu}{3\nu_0 \gamma^2}\left[1+\left(\frac{\gamma \nu_p}{\nu}\right)^2\right]^{3/2}
\end{equation}
\begin{equation}\label{eq:F}
  \begin{split}
    F(s) &\equiv s\int_s^{\infty}K_{5/3}(\xi)\mathrm{d}\xi\\
    &\approx 1.25s^{1/3}\mathrm{exp}(-s)[648+s^2]^{1/12}
  \end{split}
\end{equation}
where $K_{5/3}(\xi)$ is the modified Bessel function of order $5/3$.

The synchrotron energy spectrum per frequency, $EdN_{\nu}/dE$,  given populations of electrons and positrons each with an equilibrium density spectrum $dn_e/dE_e$ is:
\begin{equation}\label{eq:jv}
 E\frac{dN_{\nu}}{dE} = \frac{2m_{\chi}^2}{\langle \sigma v \rangle \rho_{\chi}^2}\int_{m_e}^{m_{\chi}} \left(\frac{dn_{e^-}}{dE}+\frac{dn_{e^+}}{dE}\right) P_{\nu} \mathrm{d} E
\end{equation}
This energy spectrum is inserted into Eq.~(\ref{eq:dmflux}) to find the limits on the annihilation cross section.

\subsection{Magnetic Field Model}

Synchrotron emission depends strongly on the magnetic field in the region of interest. From MHD simulations and from clusters with multiple Faraday RMs, the magnetic field appears to follow the gas density \citep{Murgia2004,Bonafede2010,Vacca2012}. The gas density is typically fit with a $\beta$ model \citep{Cavaliere1976}; we adjust the radial dependence with a free parameter $\eta$:
\begin{equation}\label{eq:Br}
  B(r) = B_0\left[\left(1+\left(\frac{r}{r_c}\right)^2\right)^{-(3/2)\beta}\right]^{\eta}
\end{equation}
where $B_0$ is the central magnetic field value and $r_c$ is the core radius. The parameters $\beta$ and $r_c$ are fit using X-ray data, and $B_0$ and $\eta$ are typically modeled based on a combination of simulations and RMs (e.g., for Coma, \citealt{Bonafede2010}).

We weight the magnetic field distribution by the dark matter density profile to yield a single parameter, an effective magnetic field $B_{eff}$ that is used in our synchrotron spectrum calculations:
\begin{equation}\label{eq:Beff}
  B_{eff} = \left[\frac{\int_0^{r_h}B(r)^2\rho_{\chi}^2 r^2 \mathrm{d}r}{\int_0^{r_h} \rho_{\chi}^2 r^2 \mathrm{d}r}\right]^{1/2}
\end{equation}
where $r_h$ is the radius of the considered emission region, either that of the observed diffuse emission region or the region used to place an upper limit. This weighting of the magnetic field yields a better estimate of the field in the regions where most of the synchrotron emission is originating from, i.e., regions with high dark matter densities.

\section{CLUSTER SAMPLE: RADIO AND MAGNETIC FIELD DATA}\label{sec:selection}

\subsection{Selection of Clusters from Radio Data}

We choose our sample based on radio, X-ray, and magnetic field data available from the literature. The clusters that produce the best limits on dark matter annihilation are nearby and do not host observable radio haloes, or host only low levels of central diffuse emission or mini-haloes. Unfortunately, there are very few published upper limits on the diffuse radio emission in clusters. \citet{Rudnick2009} published the most comprehensive list of upper limits (96$\%$ confidence limit) on Mpc-sized radio emission for bright X-ray clusters with redshifts between $0.03$ and $0.3$ at 327 MHz with the Westerbork Northern Sky Survey. Upper limits on Mpc-sized emission from clusters with $0.2<z<0.4$ were published in \citet{Venturi2008} using the Giant Metrewave Radio Telescope at 610 MHz. Since the astrophysical $J$ factor decreases with increasing distance, nearby clusters are the best candidates for dark matter detection; we therefore restrict our sample to $z<0.1$. Only a handful of clusters in \citet{Rudnick2009} have redshifts of less than $0.1$. Additionally, to facilitate comparisons, our sample of clusters largely overlaps with those considered for gamma-ray analysis \citep[e.g.,][]{Ackermann2010dm,Huang2012}.

We take $\beta$ model parameters, cluster masses, and gas fractions from the HIFLUGCS catalog of nearby, massive, bright X-ray galaxy clusters \citep{Chen2007}. Four clusters with published radio halo upper limits from \citet{Rudnick2009} are also part of the HIFLUGCS catalog. 

We also include three additional clusters in HIFLUGCS that have detected mini-haloes: Perseus, Ophiuchus, and A2029. Despite the detected radio flux from these clusters, they are strong candidates because they host cool cores, and thus have higher inferred central magnetic field strengths, and are also nearby and heavily dark matter dominated. Perseus and Ophiuchus especially are considered to be among the best candidates for dark matter searches \citep{Jeltema2009}. 
\begin{figure}
  \includegraphics[scale=0.4]{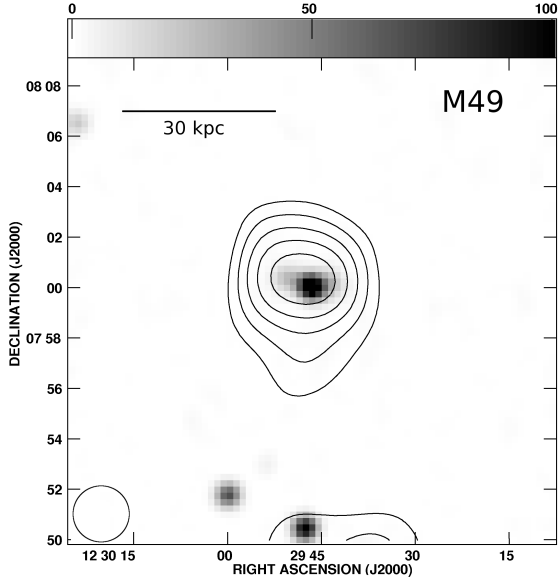}
  \caption{Greyscale image of M49 from NVSS, 45" resolution, on a scale of 0-100 mJy/45" beam. Contours show the residual diffuse emission at a resolution of 135", as shown in the lower left, after subtraction of the compact emission. Contour levels are at 2 mJy/135" beam $\times$ (1,2,3,\ldots).}
  \label{fig:M49}
\end{figure}
Coma is also included in our sample, as it has both a very well-studied radio halo \citep{Kim1990} and magnetic field \citep{Bonafede2010}, and is generally used as a representative massive, merging cluster. It is also the only cluster that has been used previously to place constraints on the dark matter content in clusters using radio data \citep{Colafrancesco2006} and is frequently considered in analyses of dark matter annihilation in clusters, especially in gamma rays (e.g., most recently in \citealt{Han2012a}). We show that it produces comparably poor limits on the dark matter annihilation cross section due to the large observed radio emission, for a smooth NFW dark matter profile with a final annihilation state of $b\overline{b}$ in Fig.~\ref{fig:NFWbb}; we find that Coma yields poor limits for other dark matter profiles and annihilation channels as well.

Finally, we searched for diffuse radio emission using the NRAO VLA Sky Survey (NVSS; \citealt{Condon1998}) at $1.4$~GHz for some of the best candidate clusters for dark matter detection which do not have published radio upper limits or detections, including Virgo, Fornax, AWM7, and two Virgo groups, M49 and NGC4636. We exclude Virgo and Fornax as they both contain very bright radio sources that likely wash out any true diffuse emission. 

For AWM7, we find no detected diffuse radio emission in the NVSS images. We consider a circular emission region with a diameter of $\sim2$ core radii, and after subtracting out point sources, we place an upper limit on the diffuse emission, following the method in \citet{Rudnick2009}. 
\begin{figure}
  \includegraphics[scale=0.4]{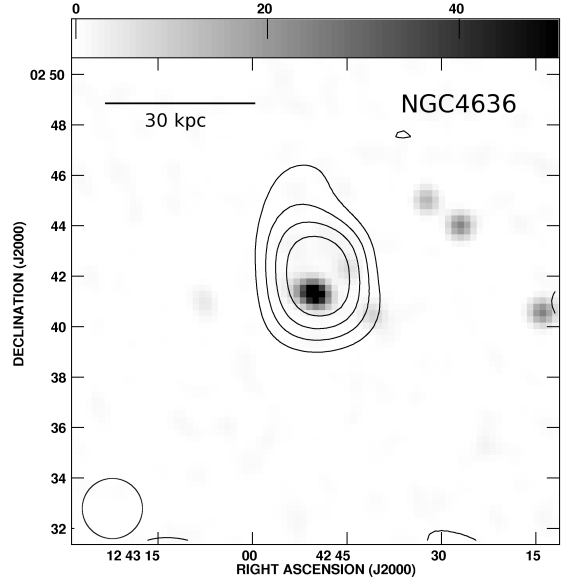}
  \caption{Greyscale image of NGC4636 from NVSS, 45" resolution, on a scale of 0-50 mJy/45" beam. Contours show the residual diffuse emission at a resolution of 135", as shown in the lower left, after subtraction of the compact emission. Contour levels are at 2 mJy/135" beam $\times$ (1,2,3,\ldots).}
  \label{fig:NGC4636}
\end{figure}

We do find extended radio emission in the NVSS from M49 and NGC4636 extending beyond the bright emission from the central galaxies, after removing the emission from compact and slightly resolved sources using a multiresolution filtering technique as used by e.g., \citet{Rudnick2009}. The resulting extended emission from M49 is approximately $\sim30-35$~kpc ($6-7\arcmin$) in diameter with a flux density of $40$~mJy, much larger and fainter than the $\sim220$~mJy emission from the central, Seyfert 2 elliptical galaxy \citep{Dunn2010,Brown2011c}. NGC4636 shows diffuse emission over a $\sim36$~kpc ($7\arcmin$) range with a flux density of $30$~mJy. The central source of NGC4636 show clear jet-like structure and is approximately $5$~kpc in size at $610$~MHz \citet{Giacintucci2011}, and this structure is not visible at $1.4$~GHz at the lower resolution of the NVSS, whereas the diffuse emission at $1.4$~GHz is larger and more spherical in shape. This diffuse central emission has not been previously identified in either group. Images of the diffuse emission for M49 and NGC4636 are shown in Figs.~\ref{fig:M49} and \ref{fig:NGC4636}, respectively.

\subsection{Magnetic Fields in Clusters}

The observation of diffuse radio emission in galaxy clusters requires the presence of large-scale magnetic fields associated with the ICM. Intracluster magnetic fields can be inferred from the Faraday rotation of polarized radiation of individual sources with typical central values of $\sim1-10~\mu$G \citep[e.g.,][]{Eilek2002,Bonafede2010,Vacca2012}. Clusters with cool cores can have higher inferred central magnetic field strengths, $\sim 10-40~\mu$G \citep{Taylor2002,Taylor2006,Kuchar2011}. Magnetic fields in clusters can also be estimated using radio halo observations assuming equipartition of cosmic ray and magnetic energy densities; these estimates are typically between $0.1-1~\mu$G and are taken as lower limits (e.g. \citealt{Govoni2004}; see also \citealt{Carilli2002} for a review of magnetic fields in clusters).

Turbulence due to past mergers can amplify magnetic fields in clusters to $\mu$G levels \citep[e.g.,][]{Dolag2002,Subramanian2006,Ryu2008}, and can also accelerate particles, which may be responsible for the giant radio haloes observed in some clusters (e.g., \citealt{Brunetti2011} and references therein). If mergers drive radio haloes, then the difference between clusters with and without haloes may be dynamical: clusters with haloes have suffered a merger in their recent past, while clusters without haloes are more dynamically relaxed, with no recent mergers in their history \citep{Brunetti2009,Cassano2010a}. Assuming clusters evolve in this way, it is natural to infer that at some point the cluster magnetic field will dissipate and suppress the synchrotron radiation producing haloes \citep{Brunetti2009,Cassano2010a}. However, theoretical studies and simulations of clusters show that large scale magnetic fields are long-lived, $\sim4-5$~Gyr \citep{Subramanian2006}, which is longer than the cosmic ray electron/positron lifetime, and thus longer than the lifetime of the radio halo, about $1$~Gyr \citep{Brunetti2009}. Additional theoretical work on the ICM by \citet{Kunz2011}, in which heating by local plasma instabilities can stabilize cooling in the ICM, predicts $\sim 1-10~\mu$G magnetic field strengths that scale weakly with gas density and generally agree with values estimated from RMs for specific clusters.

Only a handful of clusters have well-studied magnetic fields; less than half of our sample have published studies on their magnetic fields. For the clusters that are not strong cool cores as defined by \citet{Hudson2010} but do not have any published information about their magnetic fields, we use the best fit values of $B_0$ and $\eta$ for Coma from \citet{Bonafede2010}, which we call the non-cool-core model, as Coma is a well-studied non-cool-core cluster. The magnetic field in clusters may turn out to be connected to the presence of a radio halo in such a way that clusters without observed haloes have lower magnetic fields than those with haloes \citep[e.g.,][]{Brunetti2009}. However, we choose to use the magnetic field parameters derived for Coma even for clusters with upper limits on the radio emission. This is because RM studies yield no differences in the RMs, and thus inferred magnetic fields, of clusters with and without radio haloes \citep{Clarke2001,Govoni2010}.

Cool core clusters generally have higher inferred central magnetic field strengths than clusters without cool cores \citep{Kuchar2011}. In our sample both A2029 and Perseus have published central field strengths from RMs (\citealt{Eilek2002} and \citealt{Taylor2006}, respectively). A2199 is the only cool core cluster in our sample with best fit values for both $B_0$ and $\eta$ from Faraday RMs \citep{Vacca2012}. Ophiuchus, also a cool core cluster, has a published RM from a radio galaxy outside the cool core region and therefore no estimated central magnetic field strength \citep{Govoni2010}. It has one of the hottest known cool cores \citep{Fujita2008}, which could mean its central magnetic field is somewhere between that of a strong cool core cluster and a cluster with no cool core. We therefore choose the more conservative, non-cool-core model to estimate the magnetic field of Ophiuchus when comparing constraints from other clusters, but we also show how our constraints change if we use the central magnetic field value of Perseus, the prototypical cool-core cluster. We discuss the effects of the uncertainty on the magnetic field on dark matter constraints, specifically with respect to Ophiuchus and A2199, further in Sec.~\ref{sec:errana}. 

Little is known about the magnetic field strengths for groups of galaxies, especially in/near the central elliptical galaxy that typically dominates these groups. We choose to use the non-cool-core model for the magnetic field in calculating the limits for the two Virgo groups in our sample, M49 and NGC4636. These groups are heavily dominated by their central elliptical galaxies, which must host magnetic fields of their own. Large scale galactic magnetic fields of normal galaxies are typically in the $\mu$G range (e.g., for the Milky Way, \citealt{Noutsos2012}). We choose therefore to use the non-cool-core model for the magnetic fields of these groups, the central strength of which is probably comparable to or lower than the strength of the field within the central galaxy. Cluster properties, radio data, and magnetic field parameters are listed in Table~\ref{tab1}.

\section{RESULTS AND DISCUSSION}\label{sec:results}

\begin{figure}
  \includegraphics[scale=0.52]{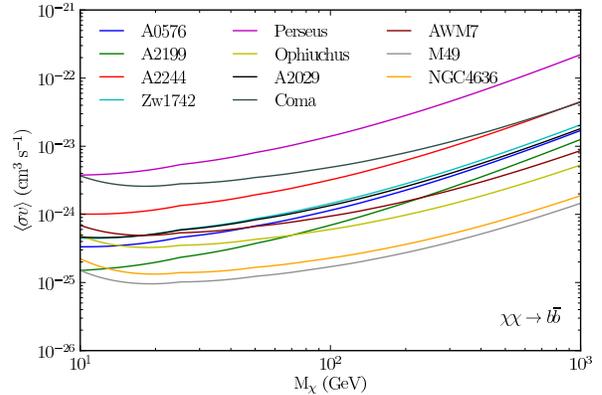}
  \caption{Dark matter annihilation cross section upper limits for all clusters in the sample; we assume a smooth NFW dark matter density profile.}
  \label{fig:NFWbb}
\end{figure}

We derive constraints on the dark matter annihilation cross section by assuming conservatively that the radio upper limits or low levels of observed emission in a sample of nearby galaxy clusters are due to synchrotron emission from electrons and positrons produced by annihilating dark matter. In Fig.~\ref{fig:NFWbb}, we present the cross section upper limits for all the clusters in our sample, each with an NFW dark matter profile, for the $b\overline{b}$ annihilation channel. For this dark matter profile, the Virgo groups NGC4636 and M49 yield the tighest constraints, while the best constraints are produced by A2199 for massive clusters with published radio data at low dark matter particle masses and by Ophiuchus at higher masses. Perseus, which hosts a particularly bright mini-halo, produces the weakest constraints. As expected, the constraints from the Coma Cluster, with its bright, giant radio halo, are also relatively poor.

\begin{figure}
  \centering
  \includegraphics[scale=0.52]{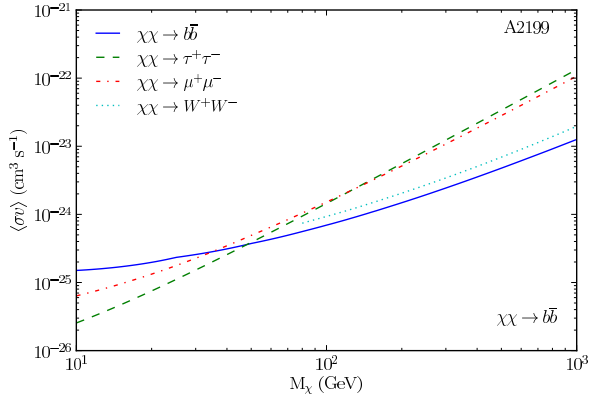}
  \caption{Dark matter annihilation cross section upper limits for A2199, for four annihilation channels, with a smooth NFW dark matter profile only.}
  \label{fig:A2199NFW}
\end{figure}

\begin{figure}
  \centering
  \includegraphics[scale=0.52]{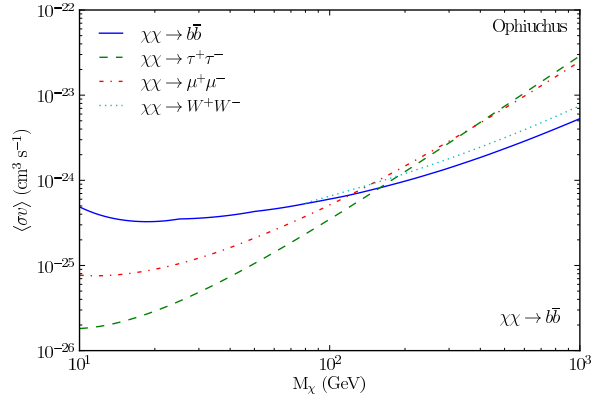}
  \caption{Dark matter annihilation cross section upper limits for Ophiuchus, for four annihilation channels, with a smooth NFW dark matter profile only. We use a central magnetic field value of $4.7\mu$G and $\eta=0.5$ (our non-cool-core model) in deriving these constraints.}
  \label{fig:OphNFW}
\end{figure}

We also consider four different annihilation channels: $b\overline{b}$, $\tau^+ \tau^-$, $\mu^+ \mu^-$, and $W^+ W^-$. In Figures~\ref{fig:A2199NFW} and~\ref{fig:OphNFW} we show the limits for each of the different channels for A2199 and Ophiuchus, using an NFW dark matter profile. The $\tau^+ \tau^-$ and $\mu^+ \mu^-$ channels, which tend to produce more electrons and positrons per annihilation than $b\overline{b}$, yield the best limits at lower particle masses, while the $b\overline{b}$ yields better limits at higher masses, which is generally true of all the clusters.

\subsection{Comparison to Limits from Gamma-Ray Emission}

Many previous studies of clusters have focused on the potential gamma-ray emission from dark matter annihilation, typically using data from \textit{Fermi} \citep{Ackermann2010dm,Ando2012,Huang2012,Nezri2012,Han2012a}. Using gamma rays to investigate the dark matter content of clusters has the advantage that the emission only depends on the underlying particle population(s) and chosen dark matter profile, and does not depend on other cluster properties, e.g., the magnetic field. However, only very nearby clusters produce useful limits, and clusters that are near the Galactic plane or contain a point source that is bright in gamma rays must be excluded from these studies, as the Galactic plane is gamma-ray bright and the resolution of \textit{Fermi} and ground-based gamma-ray telescopes is poor compared to radio telescopes. A2199 and Ophiuchus, which yield the best limits in many cases in the radio, are typically not considered for gamma-ray analysis, as A2199 is too far away, and Ophiuchus is very near the Galactic center, a very bright gamma-ray region. Comparing the limits for A2199 for dark matter annihilating to $b\overline{b}$ with a smooth NFW profile to the limits derived for Fornax, also for $b\overline{b}$ and an NFW profile, in \citet{Huang2012}, which are the best limits derived from gamma rays from a single cluster with an NFW profile to date, our limits from A2199 are approximately a factor of $\sim3$ or more across a wide range of masses.

\citet{Han2012a}, using the results of the Phoenix simulations to model substructure and considering the possibility that the gamma-ray emission from clusters is extended, placed constraints on dark matter annihilation in Coma, Virgo and Fornax that are comparable to or better than the limits derived in \citet{Huang2012} including substructure. Our limits from A2199 when considering the same substructure model from the Phoenix simulations, are comparable to those reported in \citet{Han2012a}. 

\subsection{Substructure in Clusters and Groups}

\begin{figure}
  \centering
  \includegraphics[scale=0.52]{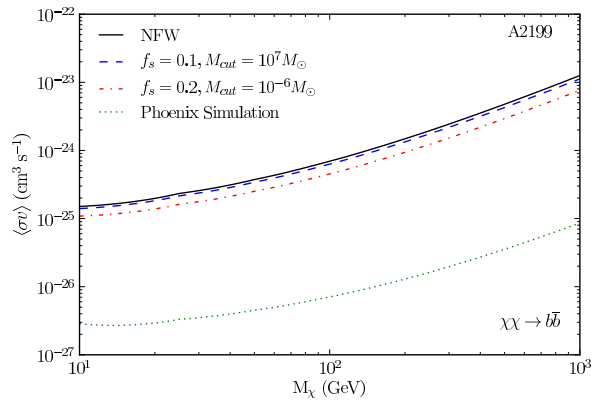}
  \caption{Dark matter annihilation cross section upper limits for A2199, for one annihilation channel ($b\overline{b}$) and our four different substructure models.}
  \label{fig:A2199bb}
\end{figure}

 The boost factors obtained by using the the Phoenix simulations are $\gtrsim10$ times larger than the boost factors derived from our other adopted substructure models, which in turn yield limits that are $1$--$2$ orders of magnitude tighter. This is reflected in Fig.~\ref{fig:A2199bb}, which shows the limits for the $b\overline{b}$ annihilation channel for our four adopted dark matter profiles for A2199. The limits from our conservative and optimistic models are only slightly better than those produced by an NFW profile, about $10-30\%$ . The limits that result from using the Phoenix profile are much lower, almost $2$ orders of magnitude for A2199, and dip below the nominal thermal annihilation cross section $3\times10^{-26}$~cm$^3$~s$^{-1}$ for masses $\lesssim400$~GeV. A similar pattern is true for most other clusters, except M49 and NGC4636. This is because the emission region is so small for these groups that the $J$ factors only change by at most a factor of four as the amount of substructure is increased, while for other clusters the difference between $J_{NFW}$ and $J_{PHX}$ is typically $2-3$ orders of magntitude.

\begin{figure}
  \centering
  \includegraphics[scale=0.52]{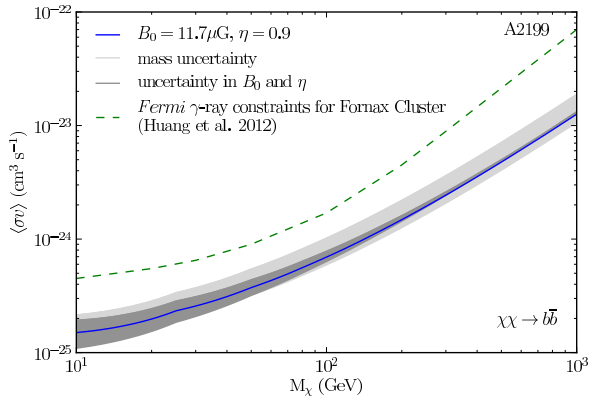}
  \caption{Effects of uncertainty in cluster mass and magnetic field parameters, for A2199, for the final state $b\overline{b}$, with an NFW dark matter profile. The constraints from the Fornax Cluster using gamma ray upper limits from \textit{Fermi} are copied from Fig.~4 of \citet{Huang2012} and also correspond to an NFW profile for final state $b\overline{b}$.}
  \label{fig:A2199error}
\end{figure}

\subsection{Uncertainties in Cluster Masses and Magnetic Fields}\label{sec:errana}

Aside from the assumed dark matter profile and amount of substructure, the two main sources of uncertainty in the limits on the dark matter annihilation cross section are the cluster mass, reported by \citet{Chen2007}, and the uncertainty in the magnetic field parameters $B_0$ and $\eta$. \citet{Chen2007} also report uncertainties in the gas fraction, the core radius, and $\beta$; we find that the uncertainties associated to these quantities are, however, negligible compared to the cluster mass and magnetic field uncertainties. 

In Fig.~\ref{fig:A2199error}, we show the relative magnitudes of the various sources of uncertainty for one of the best cluster candidates, A2199. The uncertainty in the mass leads to an uncertainty in the $J$ factors which is generally $\lesssim2$ smaller for all the clusters in our sample. For the uncertainty in the magnetic field of A2199, we choose values for $B_0$ and $\eta$ that best represent the spread in the $B_0$--$\eta$ contour plot in Figure 9 of \citet{Vacca2012}. The uncertainty in the magnetic field is comparable to the uncertainty in the cluster mass at lower dark matter masses, leading to uncertainties in the annihilation cross section of about a factor of $2$. At higher masses, the magnetic field uncertainty translates to only about a $10\%$ uncertainty in the limits while the uncertainty in cluster mass remains constant. This is because the total radio energy per frequency, and thus the annihilation cross section, depends on the magnetic field to a power proportional to the index of the underlying electron/positron distribution. The electron/positron distribution produced by lower mass dark matter particles is generally steeper than the distribution produced by higher mass dark matter particles, so the magnetic field is more important at lower dark matter masses than at higher masses.

A2199 is a cool-core cluster \citep[e.g.,][]{Hudson2010} and is therefore more likely to host a small mini-halo hundreds of kpc in size rather than a giant, Mpc-sized radio halo, as we assumed here based on the upper limits available in the literature \citep{Rudnick2009}. It is also possible that the magnetic field determined by \citet{Vacca2012} is only relevant inside the cool core region. However, a smaller emission region would likely not change our results much, since the dark matter distribution and thus emission from dark matter is highly centrally peaked (except perhaps in the case of the Phoenix simulations, where the dark matter distribution is much flatter).
\begin{figure}
  \centering
  \includegraphics[scale=0.52]{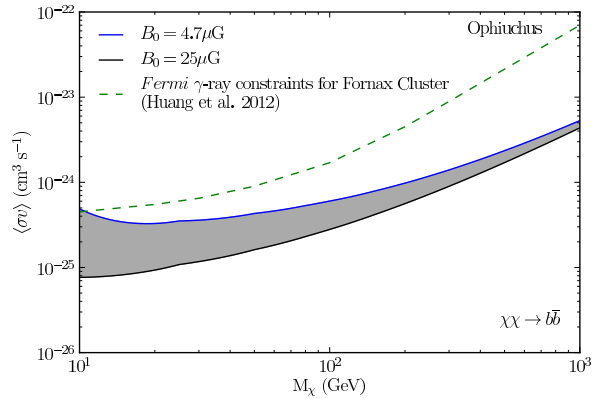}
  \caption{Effects of uncertainty in the central magnetic field strength for Ophiuchus, for one annihilation channel ($b\overline{b}$), with an NFW dark matter profile. The constraints from the Fornax Cluster using gamma ray upper limits from \textit{Fermi} are copied from Fig.~4 of \citet{Huang2012} and also correspond to an NFW profile for final state $b\overline{b}$ (same as in Fig.~\ref{fig:A2199error}).}
  \label{fig:Opherror}
\end{figure}

While Ophiuchus yields some of the best limits for the $\tau^+ \tau^-$ and $\mu^+ \mu^-$ annihilation channels, its magnetic field is not well understood. Ophiuchus has a claimed detection of nonthermal hard X-ray emission from a deep INTEGRAL observation (\citealt{Eckert2008}; see also \citealt{Profumo2008}). The magnetic field strength of Ophiuchus has been estimated to be around $0.1~\mu$G using this observation combined with radio data, assuming the hard X-ray emission is due to the IC scattering of the same population of relativistic electrons that are the source of the radio emission due to synchrotron losses \citep{Eckert2008,Perez-Torres2009,Nevalainen2009}. This is inconsistent with the larger magnetic fields inferred from Faraday RMs in similar clusters. 

Both approaches to measuring cluster magnetic fields rely on several assumptions. The field strength calculated using hard X-ray emission is typically a volume average over the size of the emission region, and it assumes that the field is uniform, while simulations point to a field that decays radially with gas density (e.g., \citealt{Donnert2009}). Additionally, assumptions must be made about the energy spectrum and origin of the cosmic ray lepton population in order to match the observed X-ray and radio emission that potentially conflict with other observations, e.g. in gamma rays \citep{Colafrancesco2009}. 

RMs rely on the gas density along the line of sight. Turbulence could distort the gas density and therefore the magnetic field along the line of sight. Additionally, it is possible that most of the contribution to an observed RM is due to the local environment around the radio source itself, implying the inferred magnetic fields may not be representative of what is happening in the ICM (\citealt{Rudnick2003}, however see also \citealt{Ensslin2003a}).

We choose to use field estimates based on RMs of other clusters for Ophiuchus, setting the non-cool-core model as the lower limit, and the field of Perseus, a cluster with a strong cool core, as the upper limit. We show the effect of the uncertainty in the magnetic field of Ophiuchus in Fig.~\ref{fig:Opherror} by varying $B_0$ with $\eta$ fixed at $0.5$. The limits vary by approximately a factor of $7$ for a mass of $10$~GeV, and a factor of $1.3$ at $1000$~GeV. 

\section{CONCLUSIONS}\label{sec:end}

We presented here new limits on the dark matter annihilation in clusters of galaxies using radio data in clusters. Overall, the limits derived from upper limits on radio haloes in massive clusters, low levels of observed central diffuse emission in galaxy groups, and detections of mini-haloes in cool core clusters are better than previously derived limits using non-detections in gamma rays for the same substructure model and annihilation channel by a factor of $\sim3$ or more. However, our limits depend strongly on both the assumed amount of substructure, which is also true for limits derived from gamma rays, and the chosen magnetic field model for any individual cluster. We consider four models for the dark matter spatial profile, with an increasing level of substructure. Using the results from the Phoenix simulations, which contain the highest fraction of substructure, the limits on the annihilation cross section are $1-2$ orders of magnitude better than those adopting an NFW profile, and lower than the thermal relic cross section for masses $\lesssim400$~GeV. The uncertainties in the magnetic field models for A2199 and Ophiuchus, which yield the best limits for a range of annihilation channels and substructure models, lead to uncertainties in the cross section limits of a factor of $2-7$ for low mass dark matter particles and a factor of $2$ or less for higher masses. We consider four dark matter pair-annihilation channels in our analysis; channels that produce harder leptons ($\tau^+ \tau^-$ and $\mu^+ \mu^-$) yield better limits at lower masses, as expected, while at higher masses, $b\overline{b}$ yields the best limits. We also report the detection of faint diffuse radio emission from M49 and NGC4636, two groups near the Virgo Cluster, using NVSS data. These groups yield very strong limits for an NFW dark matter profile, but since the emission regions are so small, they only improve by a factor of $<2$ as the amount of substructure is increased. 

\acknowledgments
This work is partly supported by NASA grant NNX11AQ10G. SP acknowledges support from an Outstanding Junior Investigator Award from the Department of Energy, and from DoE grant DE-FG02-04ER41286. At the University of Minnesota, this work is supported in part by NSF grants AST-0908688 and AST-1211595.

\bibliography{RadioDM}
\clearpage
\begin{table}\label{tab1}

\begin{center}
  Table 1\\
  Galaxy Cluster Properties
  \\
\begin{tabular}{ c c c c c c c c c c c } \hline
Name & $z$ & M$_{vir}$ & $\beta$ & $r_c$ & $\nu$ & $S_{\nu}$& R$_h$& Radio& $B_0$& $B_0$ \\ 
     &     & ($10^{14}$ M$_{\odot}$) & & (kpc) & (MHz) & (Jy) & (Mpc) & Ref. & ($\mu$G) & Ref. \\ \hline
A0576 & 0.0381 & 4.80 & 0.825 & 277 & 327 & $<$0.20 & 0.5 & 1 & ... & ... \\
A2199 & 0.0302 & 4.18 & 0.655 & 98 & 327 & $<$0.025 & 0.5 & 1 & 11.7 & 4 \\
A2244 & 0.0970 & 5.14 & 0.607 & 88 & 327 & $<$0.08 & 0.5 & 1 & ... & ... \\
Zw1742$^a$ & 0.0757 & 11.1 & 0.717 & 163 & 327 & $<$0.10 & 0.5 & 1 & ... & ...\\
Perseus & 0.0183 & 5.04 & 0.540 & 44 & 1400 & 1.979 & 0.069 & 2 & 25 & 5 \\ 
Ophiuchus & 0.0280 & 48.4 & 0.747 & 196 & 1400 & 0.1064 & 0.315 & 2 & ... & ...\\
A2029 & 0.0767 & 9.62 & 0.582 & 58 & 1400 & 0.0188 & 0.125 & 2 & 16.0 & 6 \\
Coma & 0.0232 & 9.88 & 0.654 & 241 & 1400 & 0.64 & 0.415 & 3 & 4.7 & 7\\
AWM7 & 0.0172 & 4.92 & 0.671 & 122 & 1400 & $<$0.107 & 0.122 & $^{\ast}$ & ... & ...\\
M49 & $^b$ & 0.67 & 0.592 & 7 & 1400 & 0.040 & 0.016 & $^{\ast}$ & ... & ...\\
NGC4636 & $^c$ & 0.18 & 0.491 & 4 & 1400 & 0.030 & 0.018 & $^{\ast}$ & ... & ...\\ \hline
\end{tabular}
\end{center}

\textbf{Notes.} 

$^a$ Full name for Zw1742 is ZwCl1742.1+3306.

$^b$ Distance calculated from a distance modulus of 31.15, from \citet{Villegas2010}, rather than a redshift.

$^c$ Distance calculated from a distance modulus of 31.24, from \citet{Dirsch2005}, rather than a redshift.

Cluster masses taken from \citet{Chen2007}, corrected for gas fraction and current cosmology (see text). Listed values for $z$, $\beta$ and $r_c$ also taken from \citet{Chen2007} ($r_c$ corrected for current cosmology). 

For all clusters except A2199, we assume $\eta=0.5$. We use $\eta=0.9$ for A2199, consistent with \citet{Vacca2012}. 

For clusters with no listed $B_0$, we assume $B_0=4.7\mu$G and $\eta=0.5$, i.e., our non-cool-core model. 

\textbf{References.}

$^{\ast}$ This paper. (1) \citet{Rudnick2009} (2) \citet{Murgia2009} (3) \citet{Kim1990} (4) \citet{Vacca2012} (5) \citet{Taylor2006} (6) \citet{Eilek2002} (7) \citet{Bonafede2010}

\end{table}

\end{document}